\documentstyle[12pt]{article}
\def\la{{\langle}}
\def\ra{{\rangle}}
\newcommand{\beq}{\begin{equation}}
\newcommand{\beqa}{\begin{eqnarray}}
\newcommand{\eeq}{\end{equation}}
\newcommand{\eeqa}{\end{eqnarray}}
\begin{document}

\begin{center}
{\Large \bf The time of arrival concept in quantum
mechanics\footnote{It is a pleasure to dedicate this article to
Rolf Landauer in honor of his 
70th birthday. His contributions and comments have 
stimulated our interest and motivated much of our work in    
``time in quantum mechanics''.}\vspace*{1.1cm}\\}
{\large J. G. Muga, R. Sala, 
and J. P. Palao.\vspace*{0.3cm}\\}
{\it Departamento de F\'{\i}sica Fundamental y Experimental\\
Universidad de La Laguna, La Laguna, Tenerife,
Spain\vspace*{1.2cm}\\}
{\bf Abstract}
\end{center}
\vskip 3pt
\baselineskip 12pt
The concept and the formalization of the arrival time
in quantum mechanics are discussed. Different approaches  
based on trajectories, quantization rules, time operators,
phase space techniques,
renewal equations or operational procedures are
reviewed or proposed. Open questions and loose ends are 
pointed out.     
$^{}$\vspace*{2.4cm}\\
Corresponding author: J. G. Muga\\
e-mail address: JMUGA@MOLEC1.DFIS.ULL.ES\\
Fax number: 34-22-603684\\
\newpage 
\pagestyle{plain}
\section{Introduction}
In non-relativistic mechanics    
the system or its state can be characterized by a number of 
quantities with dimension of time. 
Some, such as a resonance lifetime, are state independent,
and derived from the Hamiltonian; other characteristic times 
depend on the initial state and its dynamical evolution. Examples
are the dwell, traversal [\ref{BB}], or
arrival times, and generally the times 
required to find a dynamical variable with a given value [\ref{Busch}].
The quantization of the classical characteristic times 
is problematic and has attracted considerable attention in recent 
years [\ref{char}]. ``Traversal'' and ``arrival'' times 
have in particular been studied with many different formal
treatments.
We shall be mainly concerned with ``arrival times'' in this 
paper although the difficulties, techniques  and ideas can be easily 
translated to other characteristic times.  
\subsection{Arrival times in classical
mechanics}
Assume a structureless particle that moves in one
dimension with position $q$ and momentum $p$.
In simple cases the trajectory may cross a given spatial point $X$ 
only once [\ref{note1}],
but if a potential barrier reflects the trajectory,
or if time dependent or noisy interactions affect the motion,
the trajectory may 
cross $X$ several times. The first crossing determines the 
{\em first passage time}, and in general, the $n$-$th$ crossing
gives the the $n$-$th$ {\em passage time}. For
statistical ensembles of noninteracting particles
there is a distribution of times associated with the $n$-$th$ passage.
Provided some restrictions are imposed on the initial ensemble, the
average value and higher statistical moments can be defined from it.   
However, the distributions may exist even if the moments do not.  

Let us examine first the free motion case. 
An ensemble of freely moving noninteracting
particles will be described by a phase space
distribution $F(q,p,t)$, normalized to one, that satisfies 
$F(q,p\le 0)=0$, i.e., all particles  
move rightwards. For
free motion the trajectories cross the point $X$ only once,
and  $J(X,t)$, the probability flux or
current density at $X$, provides the distribution of
first passage or arrival times; 
$J(X,t)dt$ is the fraction of particles that
cross $X$ between $t$ and $t+dt$. Using 
\beq
J(X,t)=\int F(X,p,t) \frac{p}{m} dp\,,
\eeq   
(all integrals go from $-\infty$ to $\infty$ unless indicated
otherwise. In this particular one the lower limit can of course be
set to zero), and the trajectory equation, $q(t)=q_0+pt/m$,
the average arrival time is given for free motion by  
\beq
\int
J(X,t) t dt=\int\!\!\int F(q_0,p,0)\frac{(X-q_0)m}{p} dq_0 dp\,.
\eeq
The integral exists if the singularity at $p=0$
is cancelled, at last partially, by $F$.
An asymptotic behaviour $F\sim p^\epsilon$, $\epsilon>0$, when $p\to 0$
is sufficient. 
 
When the ensemble contains particles with momenta of arbitrary sign,    
or for more complex dynamics with multiple crossings, $J$
is not the first passage time 
distribution any more. Suppose however that 
all particles are initially on the left side and that any
particle that crosses $X$ is eliminated at the crossing instant.
When this ``absorbing boundary'' is imposed 
it is still true that the flux $J_{ab}(X,t)=-dN/dt$
provides the (unnormalized)
first passage time distribution. Here, $N=N(t)$ is the time dependent
(diminishing) ``norm'' $N=\int\!\!\int F(q,p,t) dq\,dp$, and 
\beq\label{ab}
J_{ab}(X,t)=\lim_{\epsilon\to 0}J(X-\epsilon,t),\;\;\;\epsilon>0\,.
\eeq
(Equivalently, the
particles that have crossed $X$ at least once may be labeled, 
without affecting their dynamics,
so that the first passage
time distribution becomes proportional to the flux of the
complementary subensemble of unlabeled particles.)
If not all particles are eventually absorbed, 
the distribution is normalized by dividing $J_{ab}(X,t)$ by the
total norm absorbed, $1-N(\infty)=\int dt J_{ab}(X,t)$.                  

A standard way to study first passage time distributions
in classical stochastic diffusive processes is the use of
``renewal equations'' [\ref{VK}]. 
Assume that all members of the ensemble  
have initial position $q_0=x_0$ at $t=0$. There are two ways to arrive 
at point $x<X$: Directly, via trajectories that remain in the
left subspace (with respect to $X$),
or crossing $X$ one or more times.
Let $P_0(x,x_0,t)$ be the probability density for the
``direct'' event. It can be obtained by solving the diffusion equation 
with absorbing boundary conditions at $X$. The probability of the
``indirect'' event can be decomposed 
according to the first passage time at $X$ so that 
the probability density for being at $x$ at time $t$ 
irrespective of the path may be written as    
\beq\label{re}
P(x,x_0,t)=P_0(x,x_0,t)+\int_0^t f_{X,x_0}(t') P(x,X,t-t') dt'\,.
\eeq
(A related equation was proposed by Schr\"odinger [\ref{Sch},\ref{Lum}]
in terms of cumulative probabilities.)
This is the renewal equation, where $f_{X,x_0}(t')$ is the first
passage time distribution, 
and $P(x,X,t-t')$ is the probability density for being at $x$
at time $t$ 
conditioned to having been at $X$ at time $t'$ [\ref{Mar}].
$f_{X,x_0}(t')$ is the probability flux at $X$ 
with an absorbing boundary, see (\ref{ab}), 
but it can also be obtained 
solving the integral equation by Laplace transform.
\subsection{Quantum arrival times?}
Due to the basic role of the trajectory concept to define 
``arrival'' or ``passage'', 
problems may be expected to translate the classical results 
to quantum mechanics. Allcock examined the question some time ago and
arrived at a negative conclusion [\ref{All}], but more recently his
arguments have been found
to be too pessimistic by many researchers. In particular, his claim on
the unavoidability of 
reflection by the detector has been shown to be unfounded 
[\ref{JPA},\ref{AP95}].
The very existence of difficulties makes
the time of arrival a fascinating subject where basic aspects of
quantum mechanics have  to be considered, such as the  
interpretation of the quantum formalism,
the particle-wave dualism, ambiguities in the quantization of a
classical variable, the relation between classical and quantum
mechanics, or the ``measurement problem''.
These are all difficult and not completely understood matters
so, not surprisingly, the arrival time, as well as any other 
problem concerning time in quantum mechanics is
challenging and, frequently, controversial.
Apart from the purely conceptual interest there are  
experiments that provide arrival times 
and a theoretical interpretation is required.
Indeed, general first passage time problems (where the examined
variable is not necessarily a position) are quite common. An
example is the distribution of times required for a diatomic
molecule to dissociate [\ref{VK}].  
Several complementary research avenues will be discussed in
the following sections. They  
do not necessarily conflict with each
other, since different results are generally the consequence of 
having posed different questions [\ref{Triangle}].
Work on the various routes is at different
levels of maturity and some of the many open questions are indicated. 
Because of length limitations this is not a truly exhaustive
review, so we can barely 
touch the surface of many interesting contributions.
We hope anyway that the uninitiated (and perhaps some experts) 
will benefit from the global perspective presented here.        
\section{Particle trajectories and wave features} 
Some interpretations or formulations of quantum
mechanics are based on the trajectory concept. In the ``causal''
theory of Bohm [\ref{Bohm}], or in the
stochastic interpretation [\ref{sto}], the trajectories are supposed
to be actual ones and combine 
according to standard probability rules.
(Instead, in Feynman's path integral formalism 
the contribution of each path is added as a complex
amplitude. Other formal paths
have also been explored for certain applications,
for example ``Wigner trajectories'', based on a fit of the 
dynamical equation for the Wigner function into the 
classical Liouville equation, or ``Weyl trajectories'', that retain
several useful properties in common with the 
classical ones [\ref{JCP}].) Using one of these theories definite,
and generally different
answers for characteristic times, such as the traversal time,
the dwell time and the arrival time are found.
The basic problem with these investigations 
is the uncertainty on the ultimate reality of the involved
trajectories. Are, for example, Bohm trajectories anything more than 
a possible scenario?  There is no experimental evidence to support 
or reject them as actual particle paths. Finding a way to 
answer this question would be a major breakthrough in the foundations
of quantum physics. In the ``causal'' approach to the arrival time
[\ref{Lea}], $|J(X,t)|$ (suitably normalized)
is the general arrival time distribution. The result holds for arbitrary 
$X$ even in the presence of a potential barrier.
Because all
trajectory crossings, repeated or not, are taken into account,
$|J(X,t)|$
does not correspond in general to a {\em first passage} time distribution.

Instead of relying on the problematic ``particle'' 
aspect of the quantum state, it is also possible to
pay attention to times characterizing the evolution of
the ``wave'', frequently  by
means of asymptotic methods. An example is the asymptotic
``phase time'' derived from a stationary phase argument. More detailed
information is extracted from contour deformations in the complex
plane and steepest descent methods [\ref{PRA96}].
Within this perspective, it is natural to define ``times of arrival'' 
of special wave features: the main peak, the forerunners, the centroid,
or a given percentage of probability density.
These may be very relevant when the detectors are sensitive to them, 
but should not be over-interpreted in terms of ``particle paths''.         
\section{``Operational'' procedures}
The quantum arrival time may also be handled by means of 
experiments that would provide the 
traversal time classically. Quantum theory should be able to predict 
the statistics of the experimental observations using appropriate models
for the measurement. (Inversely, it is also possible
to derive from the statistics of the recorded data the associated
quantum-mechanical observable [\ref{EW}].) 
These ``operational methods'' may provide bizarre results, possibly
opposite to classical expectations. In fact       
different, classically equivalent experiments give in 
general different outcomes.
This quantum multiplicity associated with 
a single classical quantity was already emphasized by    
Bohr [\ref{Bohr}]: ``{\it Evidence
obtained under different experimental conditions cannot be comprehended
within a single picture, but must be regarded as complementary, in the 
sense that only the totality of the phenomena exhausts the possible 
information about the objects.}''

We shall describe first two operational approaches based on  
two implementations of an ``absorbing boundary''.
\subsection{Elimination of norm at a
discretized sequence of times}
Assume that a wave packet is chopped sharply 
at a regular sequence of times $\{t_i\}$, $i=1,2,...$ separated by
$\Delta t$ in such a way that the part on the right 
of the point $X$ is eliminated [\ref{Mie},\ref{chop}]. For
a classical ensemble of noninteracting particles such a procedure
would lead in the limit $\Delta t\to 0$ to the
first arrival time distribution since the chopping is effectively   
acting as an absorbing boundary. 
However, in quantum mechanics the story is quite different.
By increasing the chopping rate a higher fraction of particles is 
reflected, and in the limit $\Delta t\to 0$ this process leads to 
total reflection.
\subsection{Arrival detectors and complex potentials}
The use of complex potentials has been proposed as a way to mimic the
classical ``absorbing boundary'' procedure to evaluate first passage
times, and to model experimental conditions
in destructive time of flight experiments  [\ref{AP95},\ref{PLA97b}].
This approach does not lead to the disappointing result of the previous
operational method. The processes taking place in the destructive
detector, such as ionizations, change the structure or internal state of
the incident particle, i.e., in scattering theory language, they change
``the channel''. Of course the full measurement can be quite complex
and involve many degrees of freedom, but 
a ``reduced'' Schr\"odinger 
equation can always be written for the incident channel in terms of a
complex (non-hermitian) potential, whose precise form  
is usually modelled phenomenologically. The norm of
the incident channel, $N$, is not conserved and may be used to define an
effective arrival time distribution, proportional to $-dN/dt$,
that will in general depend on the particular complex potential
(equivalently on the detector). Under conditions usually
met in time of flight experiments using atomic or molecular beams,  
- detector at asymptotic distance from the scattering region and
particle source - and for 
good enough absorbers $-dN/dt$ can be approximated by $J$, 
the quantum mechanical current density  {\it without}
the absorber. The difference between the first 
moments of $-dN/dt$ and $J$ (the time averages) is
the {\it dwell time} of the original particle in the complex potential
region [\ref{AP95}]. In fact the ``average'' $\int J t dt /\int J dt$
is an ideal,
apparatus-independent quantity that can also be obtained in certain 
conditions as an average of a time operator (see the discussion below).
Regarding the flux as the ideal arrival time distribution, for free 
motion, or at asymptotic distances from scatterers or sources, is an
appealing idea because of the agreement with the classical expression.  
However, a quantum mechanical state composed by positive 
momenta is compatible with a negative value of $J$ at certain times
and positions [\ref{All},\ref{BM}].
Let us first stress that this effect is
quantitatively negligible for normal practice [\ref{AP95}], so that $J$
should be a satisfactory quantity for the analysis of most 
experimental data. Nevertheless,  the fundamental objection to regard 
$J$ as a true arrival probability remains a valid one, and the
deviations of experimental distributions from $J$    
in the quantum backflow regime can and should be
studied with the aid of modern advances in experimental atomic and
optical physics.  

It would be interesting to determine if ideal 
absorbing conditions (reflection coefficients of the complex 
potential equal to zero for the wave packet momentum range)
imply a unique potential, or  any particular relation 
between $-dN/dt$ and $J$ (so that an ideal quantity could be 
defined from the operational procedure.)
Inverse scattering for real potentials in one 
dimension is a well developed field but  
very little is known about the ``inverse scattering
problem'' for complex potentials. A second open question is  
the explicit construction of models, with 
additional detector degrees of freedom,
that justify the phenomenological results on a more fundamental 
level.               
\subsection{Other measurement models}
Recently, Aharonov et al. [\ref{Ah}] have analyzed a number of 
simple, idealized ``toy models'' of arrival time measurements.
Unlike the previous approach, they explicitly include extra ``detector''
degrees of freedom in the (real) Hamiltonian. They conclude that
the free particle arrival time
cannot be measured more accurately than $\hbar/E$, where $E$ is the
kinetic energy of the particle, but it is necessary to investigate
further if this is a model dependent result or a fundamental 
limitation (they argue in favor of the later).

Schulman has proposed a theory of quantum measurement  
where the state of the studied microscopic system evolves by unitary
evolution -including generally environment and apparatus
in the Hamiltonian- to one of
the possible eigenstates of the measured observable (corresponding
to the result of the measurement) [\ref{Schulbook}].
In particular, the particle detection would require the
localization of the entire particle wave function in the detector
[\ref{SchulPLA}]. Thus this theory seemingly leads to different results
from the ones
discussed in the previous subsection, where such localization is not 
assumed. A detailed model including apparatus and environment 
would clarify the actual differences further, and the occurrence or not of 
the localization proposed; also the 
quantitative implications in the calculation of arrival times. 
Schulman has discussed experimental tests that would determine the
validity of his theory [\ref{SchulPLA},\ref{Schulbook}].

\section{Path decomposition expansion and renewal equations}
A renewal equation with the form (\ref{re}) or a related
equation in terms of cumulative probabilities could be formally 
written in the quantum case and solved by Laplace transform 
by giving some precise meaning to the symbol
$P(x_2,x_1,t_2-t_1)$ [\ref{Lum}].
This meaning however is not at all obvious or clearly defined unless
some interpretation in terms of trajectories is used where the
position becomes a Markov process. Localizing the particle around
$x_1$, e.g. with a Gaussian wave function,
is possible [\ref{Lum}], but the evolution will depend on the momentum
average and dispersion of the chosen Gaussian.   
  
A close quantum relative of the classical renewal
equation is the ``path decomposition expansion''
(PDE) of the propagator into a sum over Feynman
paths classified
according to their first passage time at $X$. For
$x, x''<X$ it takes the form
\beqa
K(x'',t''|x,0)&=&K_0(x'',t''|x,0)
\nonumber\\
&+&
\int_0^{t''}  K(x'',t''|X,t')\frac{i\hbar}{2m}\frac{d}
{dX}K_0(X,t'|x,0)\,dt'\,,
\eeqa
where $K_0$ is a restricted propagator corresponding to the
half-space $(-\infty,X)$. This expression was first derived
using Feynman path integrals [\ref{Au}], and
later by a more general operator procedure [\ref{Ha}].
An even simpler derivation follows 
from the general relation between the propagators $K$ and $K_0$,
corresponding to Hamiltonians $H$ and $H_0$, $H=H_0+V$,
\beq
K(t)=K_0(t)-\frac{i}{\hbar}\int_0^t K(t-t') V K^0(t') dt',
\eeq
by putting $H_0=\Theta(-x+X)H\Theta(-x+X)$
($\Theta$ is the Heaviside function) [\ref{unp}]. 
 
By analogy with the renewal equation, it is tempting to consider 
\beq
A\equiv\int \frac{i\hbar}{2m}\frac{d} {dX}K_0(X,t'|x,0)
\psi(x,0)\,dx
\eeq
as a  ``first passage
time amplitude'' [\ref{Ha},\ref{Yamada}].
Some caution should however be exercised  
since $K_0$ does not
correspond to an absorbing boundary but to a reflecting one,
so the analogy with (\ref{re}) is only a partial one.
Moreover, the squared modulus does not have the
correct dimensions nor will generally satisfy
``probability sum rules'' because of interferences between
paths taking different times. 
The interferences may however dissapear when coupling the particle 
with an environment. This decoherent effect and the probabilities
so obtained have been examined by Halliwell and Zafiris [\ref{HZ}]
within the ``decoherent histories approach to quantum mechanics''.
These authors point out that when decoherence is achieved the resultant
probabilities depend on the mechanism producing decoherence, and insist, 
quoting Landauer [\ref{BB},\ref{Lan}],
that time in quantum mechanics only makes sense if
the mechanism by which it is measured is fully specified.    

It is illustrative to compare the
(appropriately normalized) 
squared modulus of $A$ with the flux $J$ [\ref{unp}]. For
free motion, $K$ and the restricted propagator $K_0$
are known, and analytical results are available  
by taking as initial state, at time $t=0$,
a minimum uncertainty Gaussian wave function 
with central position and momentum $x_0$ and $p_0$, and
spatial variance $\delta^2$. If  $X=0$,   
\beq
J(0,t)=\left(\frac{2}{\pi}\right)^{1/2}
\frac{(4\delta^4 p_0m-\alpha)m\delta}
{[(t\hbar)^2+(2\delta^2 m)^2]^{3/2}}\,
e^{-\frac{2\delta^2[x_0^2m^2+2mp_0tx_0+p_0^2t^2]}
{(t\hbar)^2+(2\delta^2 m)^2}}\,,
\eeq
where $\alpha=tx_0\hbar^2$. A detailed
calculation shows that, putting $\alpha=0$, the right hand side 
is proportional, up to time independent factors, to $|A|^2$.
The shapes of $J$ and a
(normalized) $|A|^2$ are close to each other
when $2\delta^2p_0>>|x_0|\hbar.$

\section{Quantization rules and time operators}
The search for ``time operators'' and the study of their properties
has been the traditional and most popular
approach [\ref{Busch},\ref{All},\ref{Ra1}-\ref{Kobe}], 
even though 
Pauli pointed out the impossibility of a self-adjoint time operator
conjugate to a Hamiltonian with bounded spectrum [\ref{Pauli}].
There are however different ways to circumvent this objection by 
defining operators which retain at least partially the desirable
properties of a time observable. Unfortunately, the definition of
quantum operators associated with classical quantities is not
justified at present
by any fundamental quantization theory, and all known quantization rules 
are essentially heuristic recipes that may provide ambiguous, non unique,
or useless operators for some classical quantities [\ref{She}].
It is essential in each case to 
examine the properties of the operators obtained and determine their
physical content (conditions that they satisfy, domain of applicability,  
and relation to operational procedures and other quantities of
interest).           
The connection of the operators with actual measurements is 
frequently obscure: As stated by Wigner [\ref{Wi2}],  
``There is no rule that would tell us which self-adjoint operators
are truly observables, nor is there any prescription known how  
the measurements are to be carried out, what apparatus to use,
etc. In a theory with a positivistic undertone, this is a serious 
gap.''  
In principle, any operator resulting from a quantization rule 
can be associated with a property of the state of
the system, which may or may not be easily measurable or useful.   
In general, a number of conditions,
not only motivated by experiments, are imposed to 
select among the possible operators. Claims of uniqueness should then 
be taken cautiously, since they generally reflect the proponent bias
towards a group of conditions, which may not be satisfactory 
for certain purposes (and certainly not for the
sensibility of competitors!)  

Kijowski [\ref{Ki}], for the free motion case, imposing a series of
conditions compatible with the classical arrival time, and
limiting the domain to 
states with positive momentum, derived (uniquely within the stated
conditions)   
the distribution of arrival times ($X=0$)
\beq
\Pi(t;\psi(0))=\left|\frac{1}{(mh)^{1/2}}\int_0^\infty
\sqrt{p}\, e^{-ip^2t/2m\hbar}
\la p|\psi(0)\ra\,dp\right|^2\,, 
\eeq
where $\la p|\psi(0)\ra$ is the state in momentum representation at $t=0$.
Note the correct behaviour under time translation of the wave function:
$\Pi[t;\psi(0)]=\Pi[t-t';\psi(t')]$.     
Related analysis were carried out by Werner [\ref{Wer}] 
and Ludwig [\ref{Lud}], and more recently by 
several authors [\ref{Busch},\ref{GRT}-\ref{D}].
This distribution has also the remarkable property of having the same 
first moment than the flux [\ref{Ki}],
\beq\label{PiJ}
\int \Pi(t) t dt =\int J(t) t dt\,,
\eeq
and may be formally written as the square $|\la t|\psi\ra|^2$,
where $|t\ra$ is the eigenstate,
\beq
\label{eigen}
\la p|t\ra=\left(\frac{p}{mh}\right)^{1/2}e^{ip^2t/2m\hbar}\,,
\eeq
of the time operator [\ref{AB},\ref{Goto},\ref{Gia}]  
\beq
\widehat{t}=-\frac{m}{2}\left(\widehat{q}
\frac{1}{\widehat{p}}+\frac{1}{\widehat{p}}
\widehat{q}\right)\,
\eeq
(The symmetrical form
$-m\widehat{p}^{-1/2}\widehat{x}\widehat{p}^{-1/2}$    
leads in the positive momentum subspace
to the {\it same} eigenvalue equation and eigenstates as $\widehat{t}$
[\ref{Busch},\ref{GRT}].)
The normalization in (\ref{eigen}) is chosen so that 
\beq
\int  \la p|t\ra\la t|p'\ra\,dt=\delta(p-p'),\;\;\;\; p,p'>0\,.
\eeq
Dealing with states composed by coherent
combinations of momenta with arbitrary sign is more
difficult because of the  
singularity at momentum zero, but there should be a theoretical 
distribution in agreement with an experimental arrival time
distribution even if the average does
not exist. A regularization procedure has been proposed [\ref{GRT}].
The results can be very different
in this case from the ones derived from the Bohm approach
[\ref{Lea98}], and  
consideration of actual experiments in this regime would be of
much interest.           
\subsection{Phase space quantization techniques}
Quantum
states and ``observables'' can
be expressed equivalently in operator form ($\widehat{\rho}$ and
$\widehat{G}$ respectively) or by means of various phase space
representations or images [$F(q,p)$ and $g(q,p)$ respectively] 
in such a way that the expectation value of the operator 
can be written as  
\beq
\la\widehat{G}(\widehat{q},\widehat{p})\ra=
\int\!\!\int\, F(q,p)\, g(q,p)\, dq\,dp.
\eeq
Sets of four transformations    
\beq\label{map}
\widehat{\rho}\,{\stackrel{\textstyle\rightarrow}
{\leftarrow}}\,F,
\;\;\;\;
\widehat{G}\,{\stackrel{\textstyle\rightarrow}
{\leftarrow}}\,g\,,
\eeq
characterized by a kernel
function $f$ can be constructed [\ref{Cohen}-\ref{PLA97}].
Each  $f$ 
defines the quantization rule $g\to \widehat{G}$,
\beq\label{517}
\widehat{G}(\widehat{q},\widehat{p})
=\frac{1}{4\pi^{2}}\int\!\!\int\!\!\int\!\!\int
g(q,p)e^{-i(\theta q+\tau p)}f(\theta,\tau)
e^{i(\theta\widehat{q}+\tau\widehat{p})}\, dq\, dp\,
d\theta\, d\tau\,,
\eeq
and the phase space images of states and observables. The
Weyl-Wigner formalism, where the state is represented by the Wigner 
function, $F^W$, and the quantization rule is given by Weyl's
prescription, corresponds to $f=1$ [\ref{WW}].      

We shall first look at the  
free motion case.
The classical time of arrival at point $X$ when the trajectory   
starts at $q_0,p$ at time $t=0$ is given by  
\beq\label{time}
t=\frac{(X-q_0)m}{p}\,.
\eeq
Inserting (\ref{time}) in  
(\ref{517}) the corresponding operators are obtained as   
\beqa
\widehat{t}&=&\frac{X m}{2\pi}\int\!\!\int
\frac{1}{p}e^{-i\tau(p-\widehat{p})}
f(0,\tau)dp\, d\tau
\nonumber\\
&-&\frac{m}{2\pi i}\int\!\!\int\frac{1}{p}
\bigg{[}f'_{\theta=0}(\theta ,\tau)+
f(0,\tau)\bigg{(}\frac{1}{2}i\hbar \tau+
i\widehat{q}\bigg{)}\bigg{]}
e^{-i\tau(p-\widehat{p})}dp\, d\tau.
\eeqa
A wide family of quantization rules, and in particular the 
ones by Weyl, Rivier and Born-Jordan
[\ref{Cohen},\ref{PLA97}] 
lead to the same operator we have already discussed, 
expressed now for arbitrary $X$,
\beq
\widehat{t}=\frac{Xm}{\widehat{p}}-m\bigg{(}\widehat{q}\frac{1}
{\widehat{p}}
-\frac{\hbar}{2i}\frac{1}{\widehat{p}^2}\bigg{)}=
\frac{Xm}{\widehat{p}}-\frac{m}{2}\left(\widehat{q}
\frac{1}{\widehat{p}}+\frac{1}{\widehat{p}}
\widehat{q}\right)\,,
\eeq
while the standard and antistandard quantizations do not produce 
a hermitian operator so they will not be discussed further. 
Several properties of this operator are easily proved in phase space
using the Weyl-Wigner formalism, $f=1$. In particular,
the relation (\ref{PiJ}) can be derived using 
$F^W(q,p,t)=F^W(q_0,p,0)$   
 (valid for free motion, $q=q_0+tp/m$), and noting that 
the phase space representative of the flux operator is the classical
expression $\delta(X)p/m$.   
\section{Other phase space techniques}
Generalizing the free motion arrival time operator is not simple, 
and only a few interaction potentials [\ref{Goto},\ref{Kobe}] or
asymptotic distances from a
scattering potential [\ref{DM}] have been worked out. 
The classical expression for the trajectory may become very
involved, and it is rarely explicit (as a function of $q_0$, $p_0$,
and $t$), so that considering different operator orderings becomes
cumbersome or impossible in practice.   
There is the additional difficulty that the equation 
$X=q(q_0,p_0,t)$ has in general real solution only for 
a limited domain of the initial phase space.

We shall sketch here how the phase space formalism combined with the 
Heisenberg picture can be used to 
provide such a generalization, and to deal with quantities
different from position. The phase space images $g^H$ of the Heisenberg
operators $\widehat{G}(t)$ depend 
on the initial phase space point $(q_0,p_0)$ and on time, 
\beq
{\it tr}\left[\widehat\rho(0)\widehat{G}(t)\right]=
\int\!\!\int  F(q_0,p_0,0) g^H(q_0,p_0,t) dq_0 dp_0\,.
\eeq
(In particular the images of the Heisenberg operator for 
position, $x^H(q_0,p_0,t)$, are in the classical limit
classical trajectories.) Suppose that the equation for an arbitrary 
$g^H$,  
\beq
\label{GG}                
g^H(q_0,p_0,t)=\cal G\,,
\eeq
where ${\cal G}$ is a predetermined value, 
has at least one solution for $t>0$.
We can identify the sequence of times
$t^{(i)}_f(q_0,p_0,{\cal G})$, $i=,1,2,...$ 
where $g^H(q_0,p_0,t)$ ``crosses''  ${\cal G}$.
An average time (note the dependence on $f$) is then  
defined for the $i$-th crossing as
\beqa
\langle t_i \rangle&=&\frac{1}{N_i}\int\!\!\int_{D_i}  F(q_0,p_0,0)
t^{(i)}_f(q_0,p_0,{\cal G}) dq_0 p_0\,,\\
N_i&=&\int\!\!\int_{D_i}  F(q_0,p_0,0) dq_0 dp_0\,,
\eeqa
where the domain of integration is restricted to the phase space
region $D_i$ where (\ref{GG}) has an $i$-th  solution,  $N_i$ is a
normalization
constant, and $f$ may be tailored in order to satisfy consistency 
requirements [\ref{Ki}] or experimentally obtained values.
Higher moments can be obtained similarly.           
\section{Average ``presence'' times}
In several of the previous sections the probability flux $J$ has  
been emphasized as an important quantity in relation to the arrival
time. We have seen in particular that classically 
it provides the first passage distribution for free motion, or 
for a general case when absorbing boundaries are imposed. 
The probability density $\rho$ does not play this role. However,
average times can also be defined in terms of it. It is
appropriate to use a different name for them, for example
average ``presence'' times,
\beq
\widetilde{t}=\frac{\int \rho(X,t) t dt}{\int \rho(X,t) dt}
\eeq
They are of interest for detectors sensitive
to the presence of the particle rather than to the flux.  
These times, and associated operators have been studied by several
authors [\ref{Wi}-\ref{OR}]. 
\section{Concluding remarks}
Understanding the various aspects of the time of arrival in
quantum mechanics remains an exciting technical and 
conceptual challenge.  
Several theoretical approaches have been proposed or reviewed.
Experiments in non-classical regimes (with backflow,
or for motion governed by an interaction potential) would provide a
much needed reference to refine operational models and ascertain the
practical relevance of intrinsic quantities or operators.  

{\bf Acknowledgments}

We acknowledge discussions with
C. R. Leavens, V. Delgado, S. Brouard, N. Yamada and J. Le\'on.   
This work has been supported by Gobierno de Canarias (Grant PI2/95) 
and Ministerio de Educaci\'on y Ciencia (Grant PB 93-0578).

\newpage
\begin{enumerate}
\item\label{BB}
R. Landauer, Ber. Bunsenges. Phys. Chem. {\bf 95}, 404 (1991)
\item\label{Busch} P. Busch, M. Grabowski, P. J. Lahti, 
Phys. Lett. A {\bf 191}, 357 (1994)
\item\label{char} J. G. Muga, S. Brouard, V. Delgado and J. P. Palao,
in Tunnelling and its implications, D. Mugnai, A. Ranfangi,
L. S. Schulman, eds., 
World Scientific, Singapore, 1997, p. 34
\item\label{note1} Of course it is also possible not the cross
the point if the particle is trapped in a spatial region or if it is 
appropriately directed.
\item\label{VK}N. G. Van Kampen,
Stochastic Processes in Physics and Chemistry,
North-Holland, Amsterdam, 1981 
\item\label{Sch}E. Schr\"odinger, Phys. Z. {\bf 16}, 289 (1915)
\item\label{Lum}O. Lumpkin, Phys. Rev. A {\bf 51}, 2758 (1995)
\item\label{Mar}In general the functions
$P(x_2,t_2|x_1,t_1)$ will depend also 
on the momentum distribution of the initial ensemble. 
Only if the position becomes a Markovian variable (diffusive motion) 
this dependence vanish.
\item\label{All} G. R. Allcock, Ann. Phys. (N. Y.) {\bf 53}, 253
(1969); {\bf 53}, 286 (1969); {\bf 53}, 311 (1969)
\item\label{JPA} S. Brouard, D. Mac\'\i as and J. G. Muga, J. Phys. A
{\bf 27}, L439 (1994)
\item\label{AP95} J. G. Muga, S. Brouard and D. Mac\'\i as, Annals
of Physics {\bf 240}, 351 (1995)
\item\label{Triangle} A beautiful illustration of
complementary answers is provided in van Kampen's
book [\ref{VK}]: What is the probability to find a chord 
larger than $3^{1/2}$ among the chords drawn at random in a 
circle of radius 1? The answer is shown to be $1/2$, $1/3$ or $1/4$,
depending on the precise meaning one gives to the 
expression ``at random''.
\item\label{Bohm} D. Bohm and B. J. Hiley, The Undivided Universe, 
Routledge, London 1993, and references therein
\item\label{sto} E. Nelson, Phys. Rev. {\bf 150}, 1079 (1966);
L. de La Pe\~na, J. Math. Phys. {\bf 10}, 1620 (1969)
\item\label{JCP} R. Sala, S. Brouard, and J. G. Muga, J. Chem. Phys.,
{\bf 99}, 2708 (1993)
\item\label{Lea} W. R. McKinnon and C. R. Leavens, Phys. Rev. A
{\bf 51}, 2748 (1995)
\item\label{PRA96} S. Brouard and J. G. Muga,
Phys. Rev. A {\bf 54}, 3055 (1996)
\item\label{EW} B. G Englert and K. W\'odkiewicz, Phys. Rev. A
{\bf 51}, R2661 (1995)
\item\label{Bohr} N. Bohr in ``Albert Einstein: Phylosopher-Scientist'',
ed. by P. A. Schlipp, Library of the Living Phylosophers,
Evanston, 1949, p. 210 
\item\label{Mie} B. Mielnik, Found. Phys. {\bf 24}, 1113 (1994)
\item\label{chop} Let $\psi_+(t_i)$ be, the wave packet
immediately after the cut
at time $t_i$, the general mapping between two states just after 
two consecutive cuts is given in momentum representation by 
\beq
\la p|\psi_+(t_{i+1})\ra=
\frac{1}{2\pi i}\int \frac{e^{-i{p'}^2\Delta t/(2m\hbar)}
\la p'|\psi_+(t_i)\ra}{p'-p-i0}\,dp'\,.
\nonumber
\eeq        
\item\label{PLA97b} J. P. Palao, J. G. Muga, S. Brouard and A. Jadcyk, 
Phys. Lett. A {\bf 233}, 227 (1997); Ph. Blanchard and A. Jadczyk, 
Helv. Phys. Acta {\bf 69}, 613 (1996).
\item\label{BM} A. J. Bracken and G. F. Melloy, J. Phys. A
{\bf 27}, 2197 (1994)
\item\label{Ah} Y. Aharonov, J. Oppenheim, S. Popescu, B. Reznik,
and W. G. Unruh, quant-ph/9709031  
\item\label{Schulbook}  L. S. Schulman, Time's arrows and
quantum measurement, 
Cambridge U. P., Cambridge, 1997
\item\label{SchulPLA} L. S. Schulman, Phys. Lett. A {\bf 130},
194 (1988)
\item\label{Au} A. Auerbach and S. Kivelson, Nucl. Phys. B {\bf 257},
799 (1985)
\item\label{Ha} J. J. Halliwell, Phys. Lett. A {\bf 207},  237 (1995)
\item\label{unp} J. G. Muga, unpublished results
\item\label{Yamada} N. Yamada, Proc. of ``Path Integrals: Theory
and Applications'', Dubna, May 1996
\item\label{HZ} J. J. Halliwell and E. Zafiris, Phys. Rev. D, accepted
\item\label{Lan} R. Landauer, Rev. Mod. Physics {\bf 66}, 217 (1994)
\item\label{Ra1} M. Razavi, Am . J. Phys.
{\bf 35}, 955 (1967)
\item\label{Ra2} M. Razavi, Il Nuovo Cimento {\bf 63}, 271 (1969)
\item\label{Ra3} M. Razavi, Can. J. Phys. {\bf 49}, 3075 (1971)
\item\label{Ki} J. Kijowski, Rep. Math. Phys. {\bf 6}, 362 (1974)
\item\label{Wer} R. Werner, J. Math. Phys. {\bf 27}, 793 (1986)
\item\label{Lud} G. Ludwig, Foundations of Quantum Mechanics II, 
Springer-Verlag,  New York, 1985
\item\label{GRT} N. Grot, C. Rovelli and R. S. Tate, Phys. Rev.
A {\bf 54}, 4676 (1996) 
\item\label{Leon} J. Le\'on, J. Phys. A {\bf 30}, 4791 (1997)
\item\label{Gia} R. Giannitrapani, preprint, quant-ph/9611015 (1996)
\item\label{DM} V. Delgado and J. G. Muga, Phys. Rev. A, accepted
\item\label{D} V. Delgado, Phys. Rev. A, accepted
\item\label{AB}Y. Aharonov and D. Bohm, Phys. Rev. {\bf 122},
1649 (1961)
\item\label{Paul} H. Paul, Ann. Phys. (Leipzig) {\bf 9}, 252 (1962)
\item\label{Goto} T. Goto, K. Yamaguchi and N. Sudo, 
Prog. Theor. Phys. {\bf 66}, 1525 (1981)
\item\label{Kobe} D. H. Kobe, Am. J. Phys. {\bf 61}, 1031 (1993);
D. H. Kobe and V. C. Aguilera Navarro, 
Phys. Rev. A {\bf 50}, 933 (1994)
\item\label{Pauli} W. Pauli, in Encyclopedia of Physics, edited by 
S. Flugge, Vol. 5/1, Springer, Berlin, 1958, p. 60 
\item\label{She} J. R. Shewell, Am. J. Phys. {\bf 27}, 16 (1959)
\item\label{Wi2} E. P. Wigner, in Contemporary Research in the
Foundations and phylosophy of Quantum Theory, ed. C. A. Hooker
Dordrecht, Reidel, 1973,  pg. 369
\item\label{Lea98} C. R. Leavens, personal communication.
\item\label{Cohen} L. Cohen, J. Math. Phys. {\bf 7}, 781 (1966) 
\item\label{PLA97} R. Sala, J. G. Muga and J. P. Palao, Phys. Lett. A
{\bf 231}, 304 (1997)
\item\label{WW} E. Wigner, Phys. Rev. A {\bf 40}, 749 (1932);
H. Weyl, The Theory of Groups and Quantum
Mechanics, Dover, New York, 1950
\item\label{Wi} E. P. Wigner, in ``Aspects of Quantum Theory'', ed. by
A. Salam and E. P. Wigner, Cambridge, London, 1972, p. 237
\item\label{BaMe}M. Bauer and P. A. Mello, Ann. Phys. (NY) {\bf 111},
38 (1978)
\item\label{Mel}M. Bauer, P. A. Mello and K. W. Mc Voy,
Z. Physik A {\bf 293}, 151 (1979)
\item\label{OR} V. S. Olkhovsky, E. Recami and A. J. Gerasimchuk,
Il Nuovo Cimento {\bf 22} (1974) 263;
E. Recami in ``The Uncertainty Principle and Foundations of 
Quantum Mechanics'', ed. by W. C. Price and S. S. Cgissick eds., 
Wiley, London, 1977, p. 21 
\end{enumerate} 
\end{document}